\documentclass[aps,twocolumn,showpacs,preprintnumbers,prl,amsmath,amssymb,amsfonts,superscriptaddress,floatfix]{revtex4-1}

\usepackage[latin1]{inputenc}
\usepackage{graphics}
\usepackage{color}
\usepackage{amssymb}
\usepackage{amsmath}
\usepackage{overpic}
\usepackage{epstopdf}
\usepackage{bm}
\usepackage{graphicx}
\usepackage{subfigure}

\begin{document}

\title{K$_2$Mo$_3$As$_3$ is a Multi-Gap Electron-Phonon Superconductor}

\author{Bing-Hua Lei}
\affiliation{Department of Physics and Astronomy, University of Missouri,
Columbia, MO 65211, USA}

\author{David J. Singh}
 \email{singhdj@missouri.edu}
\affiliation{Department of Physics and Astronomy, University of Missouri,
Columbia, MO 65211, USA}
\affiliation{Department of Chemistry, University of Missouri,
Columbia, MO 65211, USA}

\date{\today}

\begin{abstract}
We show using density functional calculations that quasi-one-dimensional
K$_2$Mo$_3$As$_3$, which is closely
related to the K$_2$Cr$_3$As$_3$ and has very similar superconducting properties,
is not close to magnetism and has conventional $s$-wave electron-phonon superconductivity.
This superconductivity is of multi-gap character due to different coupling on different Fermi surface
sheets.
This is discussed in relation to the properties of this family of quasi-one-dimensional
pnictide superconductors.
The results show that this family of superconductors provides a unique opportunity for
studying the interplay of spin-fluctuations and electron-phonon superconductivity in
transition metal pnictides and offer
a path for sorting out the different proposed superconducting scenarios in this
fascinating family of pnictide superconductors.
\end{abstract}

\maketitle

The discovery of Fe-pnictide superconductors provided a new window into unconventional
high temperature superconductivity.
\cite{kamihara}
These compounds, similar to cuprates, exhibit nearness to magnetism, layered structures with
square planar transition element sheets and unconventional order parameters. 
\cite{johnston,mazin-spm}
However, these are unlike cuprates in that the order parameter is different as is the bonding
arrangement. Importantly, the magnetism and nature of electron
correlations both appear to be different.
\cite{dai,qazilbash}
More generally, progress in understanding superconductivity, especially unconventional superconductivity,
has been driven by the discovery of new classes of superconducting materials.
These include heavy Fermions, cuprates, Fe-based superconductors and the 4d transition metal oxide,
Sr$_2$RuO$_4$,
\cite{maeno,pustagow}
all of which are superconductors near magnetism, but otherwise show quite different properties.
Each of these has provided insights into the interplay between magnetism and superconductivity,
which although long studied,
\cite{berk}
remains one of the key challenges in developing understanding of
unconventional superconductivity.
\cite{scalapino}

The $A_2T_3$As$_3$, where $A$ is an alkali metal are a class of pnictide superconductors,
\cite{bao-K2Cr3As3}
with chemical similarity to the iron-pnictide superconductors,
\cite{kamihara}
which have high critical temperatures and may
be examples of spin-fluctuation mediated superconductors.
\cite{mazin-spm,scalapino,kuroki}
The $A_2T_3$As$_3$ materials show signatures suggesting unconventional superconductivity.
They differ from the Fe-based superconductors in that the structures are based on one dimensional
transition metal tubes, coordinated by As.
This is important, for example, because spin wave behavior and spin fluctuations generally depend strongly
on dimensionality.
Additionally, these materials are examples of non-centrosymmetric superconductors.
Various pairing states have been proposed, including triplet states and recently topological
superconducting behavior in connection with $p$-wave states was suggested.
\cite{liu}
Consistent with unconventional superconductivity, the NMR Hebel-Slichter peak is absent,
\cite{luo,zhi}
the upper critical fields exceed the Pauli limit,
and, at least in the Cr compounds,
the renormalization of the electronic density of states inferred from specific heat measurements is high.
\cite{bao-K2Cr3As3,kong,shao}

Superconductivity with similar $T_c$ occurs for both $T$=Cr and $T$=Mo,
with the Mo compounds showing somewhat higher $T_c$ in general.
\cite{mu-K2Mo3As3,zhao-Rb2Mo3As3,zhao-Cs2Mo3As3}
Specifically, the experimental $T_c$ = 10.4 K, 10.5 K and 11.5 K,
for K$_2$Mo$_3$As$_3$, Rb$_2$Mo$_3$As$_3$ and Cs$_2$Mo$_3$As$_3$, respectively,
as compared to 6.1 K, 4.8 K and 2.2 K for the corresponding Cr compounds.
\cite{bao-K2Cr3As3,tang-Rb2Cr3As3,tang-Cs2Cr3As3}
We note that persistence of superconductivity under substitution with a 4$d$ element
does not preclude spin-fluctuation induced unconventional superconductivity,
as seen for example in the nearness to magnetism of Sr$_2$RuO$_4$,
and the fact that Fe-based superconductivity persists up to $\sim$35\%
Ru substitution for Fe in (Ba,Sr)(Fe,Ru)$_2$As$_2$, and in fact is induced by Ru alloying starting with
stoichiometric BaFe$_2$As$_2$.
\cite{schnelle,sharma}
Magnetism is less common in Mo compounds compared to Cr, but Mo moments are important
in materials for example in the room temperature ferromagnetism of double perovskite Sr$_2$FeMoO$_6$, \cite{kobayashi}
provide magnetism in other materials as well,
\cite{guguchia,hagmann}
and has been suggested as important in K$_2$Mo$_3$As$_3$.
\cite{bao-K2Cr3As3,tang-Rb2Cr3As3,tang-Cs2Cr3As3}
In any case, substitution of Mo for Cr provides a mechanism for tuning the magnetism, which may provide
insights in the superconductivity, especially considering similarities between these
such as the anomalously high upper critical fields in both Mo and Cr compounds.

The $A_2$Cr$_3$As$_3$ superconductors show multi-sheet Fermi surfaces, with both one-dimensional and three
dimensional sections
\cite{wu-mag,jiang-dft,subedi,zhou,zhang-spm}
and clear signatures of magnetism both from first principles calculations
\cite{wu-mag,jiang-dft,zhang-spm}
and experiments. \cite{adroja}
These Cr compounds are close to ferromagnetism as well as antiferromagnetism.
This is seen in $^{75}$As nuclear quadrupole resonance (NQR) measurements, for example,
where it was found that substitutions on the alkali metal site tune the proximity to a
ferromagnetic quantum critical point as well as the superconducting critical temperature.
\cite{yang,luo}
Ferromagnetic fluctuations are generally destructive to standard $s$-wave
superconductivity, \cite{berk}
and may favor triplet superconductivity,
\cite{wu-mag,wu-triplet}
analogous to proposals for Sr$_2$RuO$_4$.
\cite{rice,mazin-Sr2RuO4}
Neutron scattering measurements, which are more sensitive to antiferromagnetism, find spin fluctuations near
(0,0,1/2),
indicating a competition between ferromagnetic and antiferromagnetic states in this system.
\cite{taddei-mag}

However, the behavior of the Cr compounds is complicated by a structural instability of the Cr$_3$As$_3$
wires. This does not fully order, is affected by stoichiometry, including
a tendency to take up H when K deficient, and does affect the magnetic, electronic and
superconducting behavior.
\cite{taddei-233,xing-KCr3As3,taddei-H,wu-H}
This leads to wide variety of proposed electronic and superconducting states,
including evidence for Tomonaga-Luttinger liquid physics, \cite{watson}
spin-triplet superconductivity,
\cite{wu-triplet,zhong}
topological superconductivity
\cite{liu,xu}
and nodal gap functions.
\cite{shao,adroja,adroja-2,yang}

Here, we use first principles calculations to address the properties of K$_2$Mo$_3$As$_3$.
This provides a unique opportunity for sorting out the various proposals for superconductivity in this
family of compounds.
We find that K$_2$Mo$_3$As$_3$ is very different from K$_2$Cr$_3$As$_3$, and in particular unlike that compound,
K$_2$Mo$_3$As$_3$ does not show phonon instabilities in the ideal hexagonal structure.
This facilitates analysis of the magnetic and electronic properties. We find that the electronic
structure is similar to that of K$_2$Cr$_3$As$_3$, with both 1D and 3D Fermi surface sheets.
However, the magnetism is very different. We do not find proximity to ferromagnetism, nor do we
find any ordered antiferromagnetic phase.
The stability of the ideal structure enables electron phonon calculations.
We find remarkably that the superconductivity of
K$_2$Mo$_3$As$_3$ can be well described
by standard electron-phonon theory,
and additionally that it is a multi-gap superconductor providing an explanation for the
high observed upper critical field.

\begin{figure}[tbp]
\centerline{\includegraphics[width=\columnwidth]{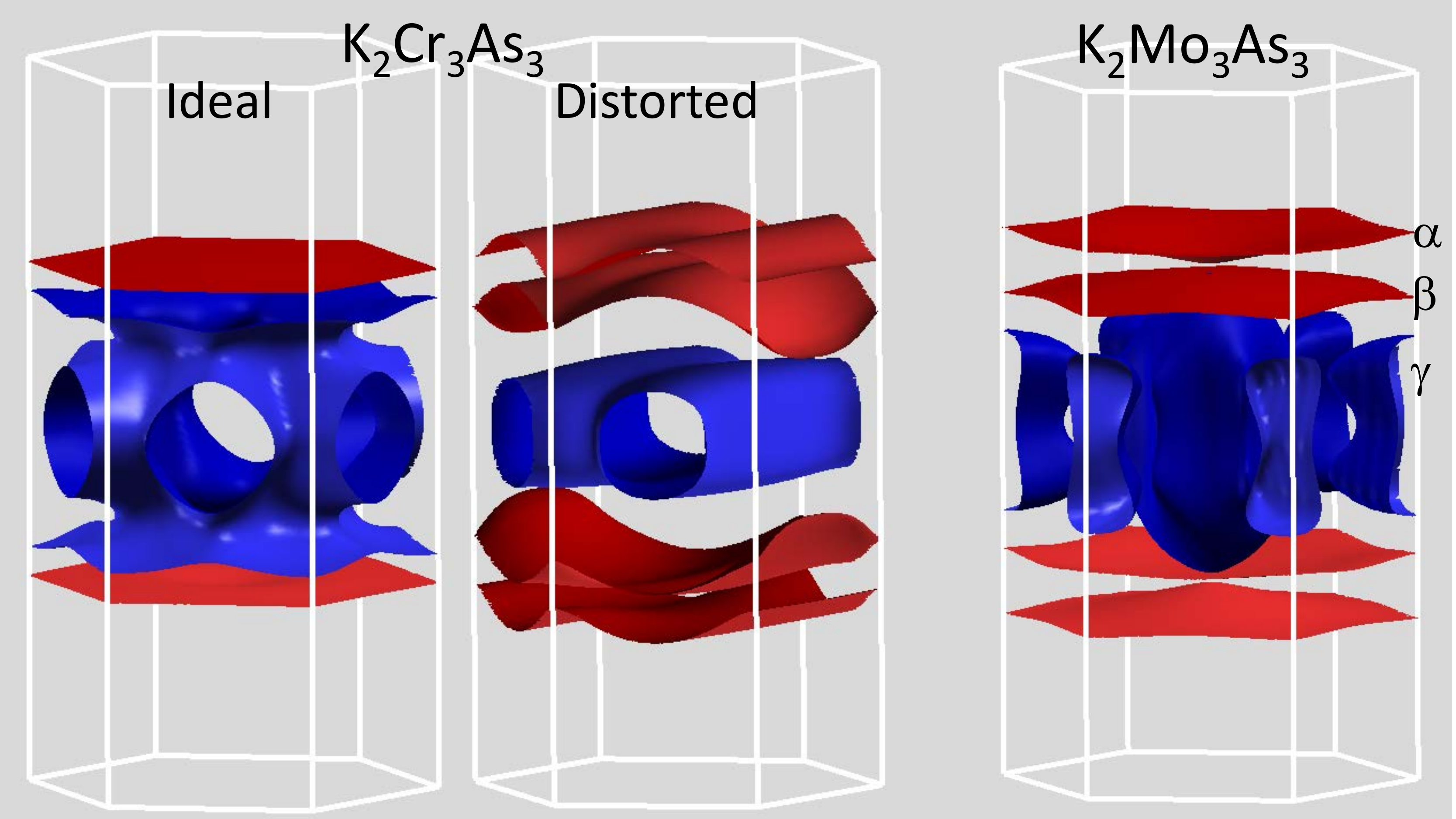}}
\caption{Fermi surfaces of ideal and distorted K$_2$Cr$_3$As$_3$ and K$_2$As$_3$Mo$_3$ as obtained
with the PBE-GGA functional. The labels of the Fermi surfaces for K$_2$Mo$_3$As$_3$ are as shown.}
\label{fermi}
\end{figure}

\begin{figure}[tbp]
\centerline{\includegraphics[width=\columnwidth]{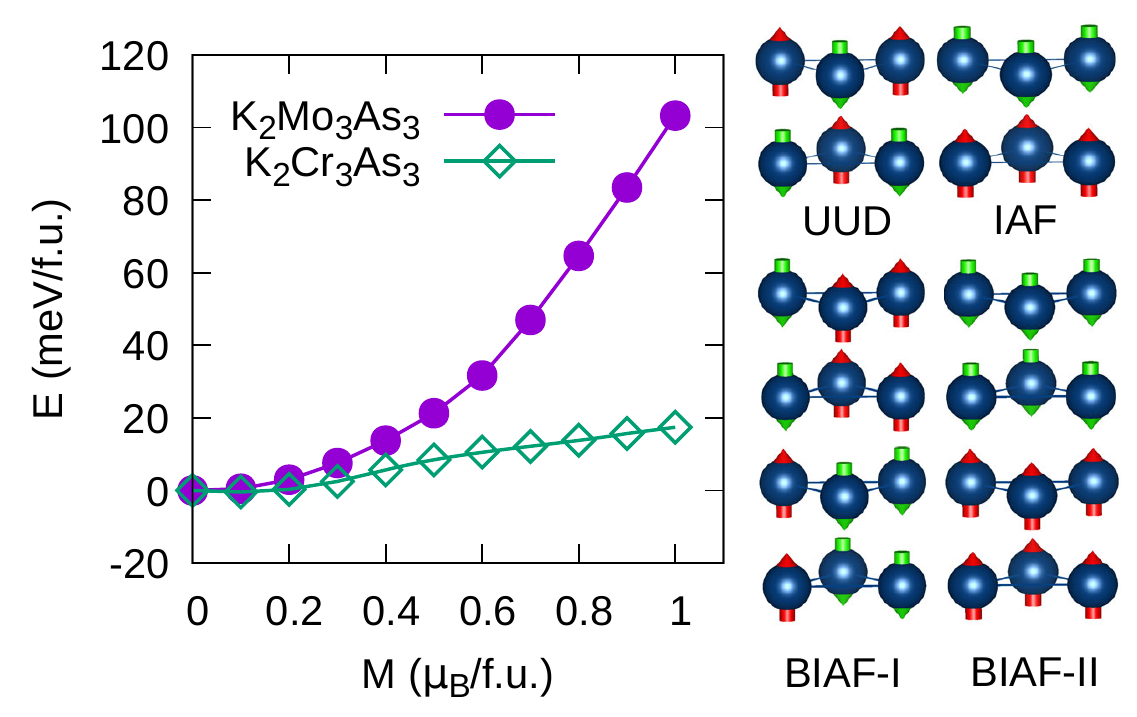}}
\caption{Comparison of fixed spin moment energy as a function of spin magnetization for K$_2$Mo$_3$As$_3$
and K$_2$Cr$_3$As$_3$,
obtained with the reported crystal structures and relaxed internal atomic coordinates (left panel);
collinear magnetic orders investigated showing the Mo atoms comprising a $c$-axis chain (right panel).}
\label{fsm}
\end{figure}

% \begin{figure}[tbp]
% \centerline{\includegraphics[width=0.7\columnwidth]{mag.pdf}}
% \caption{Collinear magnetic orders (IAF, UUD and BIAF) investigated, showing the Mo atoms comprising a $c$-axis chain.}
% \label{mag-order}
% \end{figure}

We did density functional theory (DFT) calculations with the 
Perdew, Burke, Ernzerhof generalized gradient approximation (PBE-GGA),
\cite{pbe}
and the general potential linearized augmented planewave method as implemented in the WIEN2k code.
\cite{wien2k}
The lattice parameters of the fully relaxed structure are $a$=10.256 \AA, $c$=4.463 \AA,
which are in good accord with the reported experimental values, $a$=10.145 \AA, $c$=4.453 \AA.
\cite{zhao-Rb2Mo3As3}
Emphasizing the connection of the Cr and Mo based-compounds, we note
that they show remarkably similar electronic structures as seen in the Fermi surfaces of
Fig. \ref{fermi}.
Ideal structure K$_2$Cr$_3$As$_3$ shows two almost degenerate 1D Fermi surface sheets and a 3D sheet.
The 3D sheet distorts and the near degeneracy of the two 1D sheets is lifted with the structure distortion.
K$_2$Mo$_3$As$_3$ similarly shows a 3D sheet and two 1D sheets, with changes in detail particularly a 
lifting of the near degeneracy of the 1D sheets relative to K$_2$Cr$_3$As$_3$.
The fractional band occupancies are 0.910 ($\gamma$), 0.609 ($\beta$) and 0.481 ($\alpha$), for the 3D and the
two 1D sheets in K$_2$Mo$_3$As$_3$, as compared to
0.868, 0.572 and 0.560 for ideal K$_2$Cr$_3$As$_3$ and
0.901, 0.615 and 0.484 for distorted K$_2$Cr$_3$As$_3$.
The electronic density of states at the Fermi level for K$_2$Mo$_3$As$_3$ is $N(E_F)$=4.95 eV$^{-1}$/f.u.,
with strong Mo d contributions, similar to be behavior of K$_2$Cr$_3$As$_3$, as noted previously. \cite{yang}

Turning to magnetism, the
value of $N(E_F)$ is too small to lead to a Stoner instability towards ferromagnetism.
With a reasonable Stoner parameter
\cite{sasioglu}
of $\sim$0.75 eV for 4$d$ Mo and three Mo per formula unit,
a Stoner instability would require  $N(E_F)\ge 8$ eV$^{-1}$/f.u., placing K$_2$Mo$_3$As$_3$ far from ferromagnetism.
Fig. \ref{fsm} compares fixed spin moment calculations
for K$_2$Mo$_3$As$_3$ with hexagonal
K$_2$Cr$_3$As$_3$. K$_2$Cr$_3$As$_3$ shows a weak ferromagnetic instability, consistent with
experiments showing proximity to ferromagnetism.
\cite{luo,xu}
In contrast, K$_2$Mo$_3$As$_3$ does not show any such instability.
We also find an instability for distorted K$_2$Cr$_3$As$_3$, but this is of ferrimagnetic
nature where different Cr atoms in the reduced symmetry cell take different moments, mixing ferromagnetism
and antiferromagnetism. Tests for various possible collinear antiferromagnetic arrangements for K$_2$Mo$_3$As$_3$
as shown in the right panel of Fig. \ref{fsm} show no magnetic instabilities, as anticipated from the Stoner picture.
While the aggregate Fermi surface, which defines the low energy electronic structure is three dimensional due to
the $\gamma$ sheet, there are one dimensional characteristics. True one dimensional correlated
electronic systems can show instabilities particularly charge density waves, manifested by phonon
instabilities, spin density waves, manifested by magnetic ordering, and also instability of the Fermi liquid
state.
\cite{haldane,voit}
As discussed below, we do not find phonon instabilities.
Also importantly,
the magnetic configurations tested include two (BIAF-I and BIAF-II) with a doubled unit cell along the $c$-axis,
which corresponds closely to the expected nesting vector of the most 1D $\alpha$ Fermi surface
sheet, which as mentioned has a filling of 0.481, close to 0.5.
Thus K$_2$Mo$_3$As$_3$ is much further from magnetism than the Cr compounds.

Phonons, the Eliashberg function, $\alpha^2F(\omega)$, and related parameters needed for $T_c$ were 
obtained using the Quantum Espresso and EPW codes,
with the PBE-GGA and projector augmented wave pseudopotentials
with non-linear core corrections. \cite{epw-paw}
It is essential to have reliable phonon dispersions for this purpose.
We used the EPW code \cite{epw2,epw3,epw}
to extract the Eliashberg function, $\alpha^2F(\omega)$ and band dependent gap values on the Fermi surface.
We checked these by comparing the
phonon dispersions obtained in Quantum Espresso with those obtained by supercell calculations using
the VASP code using different {\bf q}-meshes and energy cutoffs.
These calculations showed stable phonons for K$_2$Mo$_3$As$_3$,
in contrast to calculations for K$_2$Cr$_3$As$_3$.
\cite{taddei-233}
As a further test, we relaxed the crystal structure starting with various low symmetry
displacements of atoms in the unit cell. However, in all cases the structure relaxed back to the
high symmetry hexagonal structure, also in contrast to the behavior of K$_2$Cr$_3$As$_3$.
The relaxed lattice parameters for K$_2$Mo$_3$As$_3$ are $a$=10.256 \AA, $c$=4.46 \AA.
The EPW calculations were done with a Wannier interpolated
fine 36$\times$36$\times$48 {\bf k}-point grid
based on data from a 18$\times$18$\times$24 {\bf q}-grid.
Convergence was tested by using different mesh sizes.
The phonons were as obtained with Quantum Espresso. The Quantum Espresso force constants were used
in the EPW calculation, so the phonons for these calculations are the same.

\begin{figure}[tbp]
\centerline{\includegraphics[width=\columnwidth]{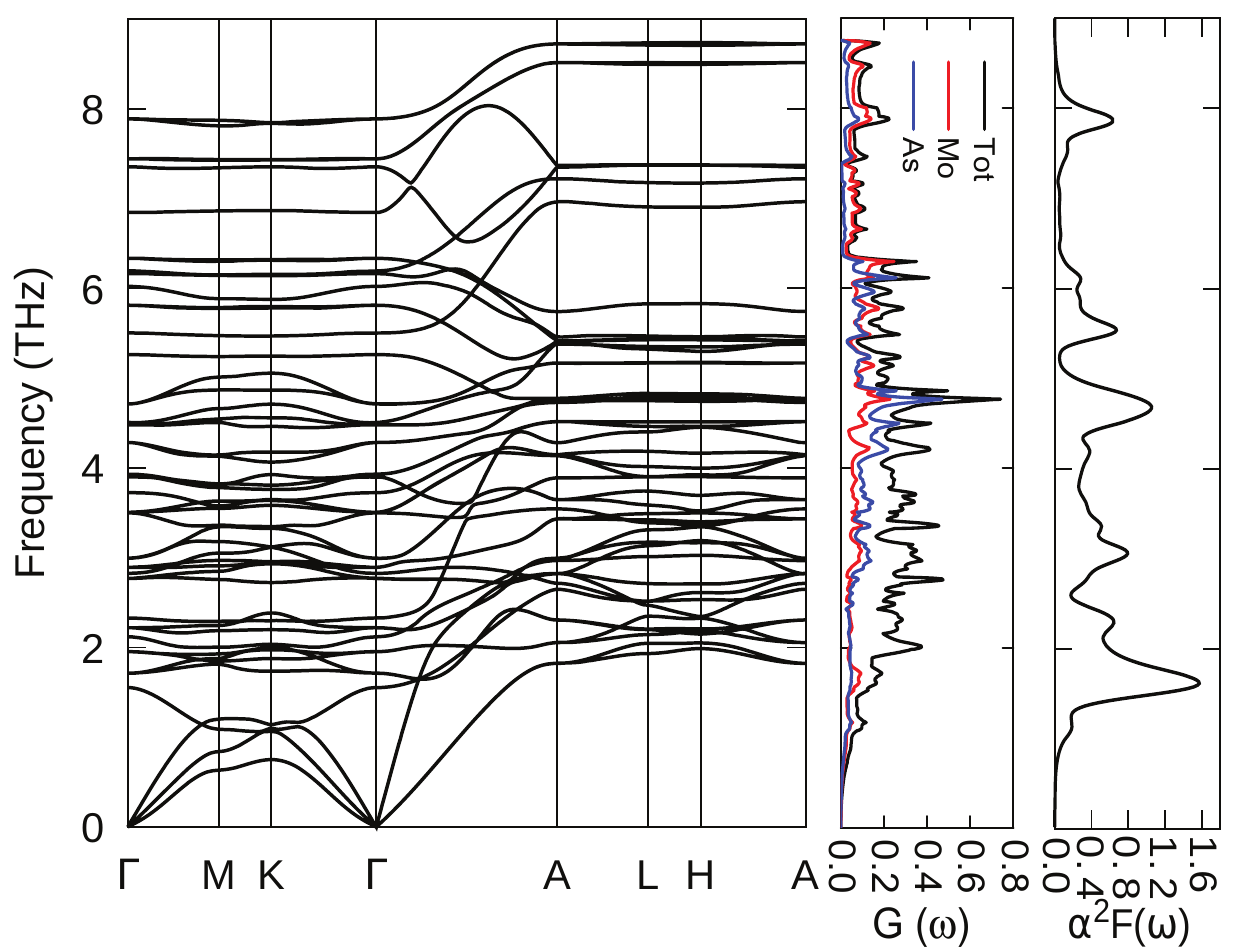}}
\caption{Calculated phonon dispersions of K$_2$Mo$_3$As$_3$, (left) the phonon density of states, 
$G(\omega)$,
with projections of Mo and As character (middle) and calculated electron-phonon $\alpha^2F(\omega)$.
A smoothing was applied to the density of states and electron phonon curves.}
\label{phonon}
\end{figure}

The phonon dispersion, phonon density of states, $G(\omega)$
and Eliashberg spectral function, $\alpha^2F(\omega)$
are given in Fig. \ref{phonon}.
The integrated electron-phonon coupling constant $\lambda=2\int(\alpha^2F(\omega)/\omega) d\omega$ is
$\lambda$=1.92.
This yields a moderate mass renormalization of $m^*/m$=$(1+\lambda)$=2.92 and
puts K$_2$Mo$_3$As$_3$ in the strong coupling regime for electron-phonon superconductivity.
The logarithmically averaged phonon frequency,
$\omega_{log}$=${\rm exp}\left((2/\lambda)\int(\alpha^2F(\omega)/\omega){\rm ln}\omega d\omega \right)$
is $\omega_{log}$=2.56 THz=123 K.
The superconducting critical temperature can be obtained using the McMillan formula as modified by
Allen and Dynes.
\cite{allen}

\begin{equation}
\label{eq:eq2}
T_c=\frac{\omega_{log}}{1.2}{\rm exp}\left[\frac{-1.04(1+\lambda)}{\lambda(1-0.62\mu^*)-\mu^*}\right],
\end{equation}

\noindent This yields predicted critical temperatures of 16.4 K and 15.2 K using reasonable values
of the Coulomb repulsion parameter, $\mu^*$ = 0.12 and 0.15, respectively.
These values are somewhat higher than the experimental value of 10.4 K, perhaps reflecting
scattering due to spin fluctuations.

\begin{figure}[tbp]
\centerline{\includegraphics[width=\columnwidth]{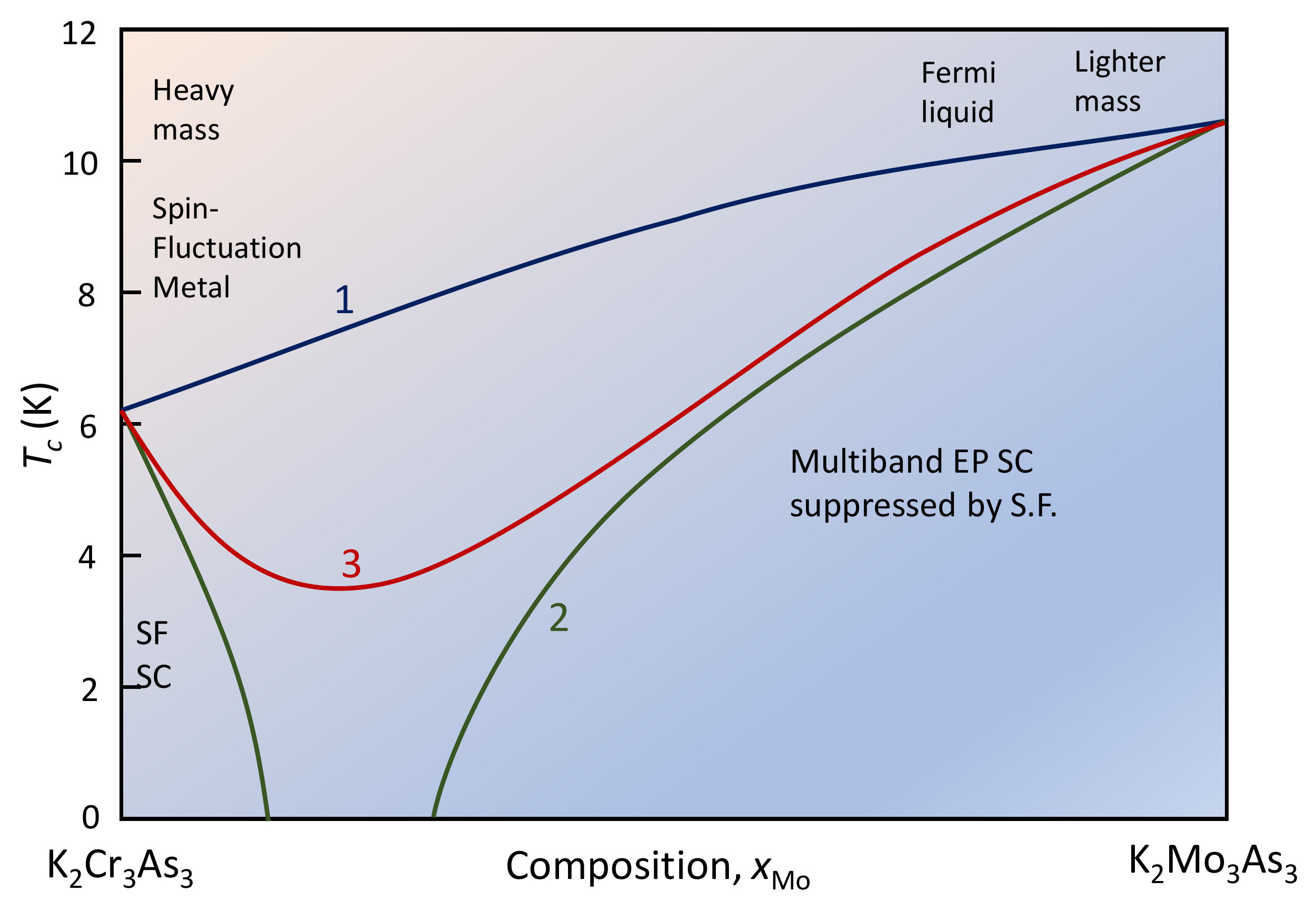}}
\caption{Three scenarios for superconductivity (see text) and
the phase diagram of the K$_2$Cr$_3$As$_3$ -- K$_2$Mo$_3$As$_3$ alloy system
that are compatible with the present results and the existing experimental data.}
\label{phase}
\end{figure}

It may be noted that there is a sizable peak in $\alpha^2F(\omega)$ at $\sim$1.8 THz.
This comes from phonons with primarily Mo character that modulate the Mo-Mo distances in the tubes.
In addition, there are significant contributions for higher frequency modes extending to the
top of the phonon spectrum. The resulting $\omega_{log}$=2.6 THz reflects these higher frequency
contributions, which contribute to the $T_c$.

Thus, K$_2$Cr$_3$As$_3$ is well described as an electron-phonon superconductor that is not in
proximity to magnetism, but that does have a multi-sheet Fermi surface with 1D and 3D sections.
As mentioned, experiments show upper critical fields, $H_{c2}$ that significantly exceed the Pauli
limit in these materials.
This is characteristic of two gap superconductivity within a standard electron-phonon scenario.
\cite{gurevich,gurevich2,gurevich3,hunte,zehetmayer}
We calculated the gap on the different Fermi surface sheets (see Fig. \ref{fermi}) using EPW.
The resulting average gap values are
$\Delta_\alpha$=2.3 meV, $\Delta_\beta$=4.8 meV and $\Delta_\gamma$=4.0 meV,
for the $\alpha$, $\beta$ and $\gamma$ sheets, respectively.
Thus, there is a considerable gap anisotropy, with much weaker pairing on the isolated 1D $\alpha$
sheet as compared to the 3D $\gamma$ sheet and the 1D $\beta$ sheet near it.
The gap values $\Delta$ are significantly higher than the weak coupling BCS value
$\Delta_{BCS}$=1.764$k_BT_c$.
This reflects strong coupling mentioned above.
These results show a multi-gap superconducting scenario, which explains
the high upper critical fields seen in this compound.

This leaves three scenarios for $A_2T_3$As$_3$ superconductivity that are compatible with experiments
and the above results.
These can be distinguished based on the phase diagram, especially the properties of the
alloy, K$_2$(Cr$_{1-x}$Mo$_x$)$_3$As$_3$ as shown schematically in Fig. \ref{phase}.
This connects the endpoint compounds - K$_2$Mo$_3$As$_3$, which
as shown above is a multi-gap electron-phonon superconductor that is
based on a Fermi liquid away from magnetism,
and K$_2$Cr$_3$As$_3$, which is a superconductor with strong signatures of spin-fluctuations.

\underline{Scenario 1:}
Both are electron-phonon superconductors; spin-fluctuations in K$_2$Cr$_3$As$_3$ suppress
$T_c$, while enhancing the effective mass and electron-scattering. Defining $\lambda_{ep}$ as the
electron phonon-coupling, and $\lambda_{sf}$ as the analogous coupling to mostly ferromagnetic and therefore
pair breaking spin-fluctuations, the
simplest model has the coupling for $s$-wave superconductivity as $\lambda_{sc}$=$\lambda_{ep}$-$\lambda_{sf}$
and specific heat mass enhancement, $\gamma/\gamma_{bare}$=1+$\lambda_{dos}$=1+$\lambda_{ep}$+$\lambda_{sf}$;
the value of $T_c$ is expected to increase monotonically with Mo content in K$_2$(Cr$_{1-x}$Mo$_x$)$_3$As$_3$,
while renormalization and other signatures of magnetism decrease. H$_{c2}$ exceeding the Pauli limit
would exist across the phase diagram reflecting multi-gap superconductivity.

\underline{Scenario 2:}
K$_2$Cr$_3$As$_3$ is an unconventional superconductor with spin-fluctuation pairing; in this case superconductivity
may be sensitive to paramagnetic impurity scattering and the proximity to the magnetic critical point; $T_c$ is
expected to decrease with Mo alloying and vanish;
emergence of $s$-wave electron-phonon mediated superconductivity is
at a higher concentration; the non-superconducting region in between would be analogous to Pd metal, which
would be a superconductor without the destructive effect of spin-fluctuations.
\cite{berk,bennemann}

\underline{Scenario 3:} That there is a cross-over regime between unconventional spin-fluctuation paired
K$_2$Cr$_3$As$_3$ and $s$-wave electron-phonon paired K$_2$Mo$_3$As$_3$; such a cross-over could be between
symmetry compatible superconducting states, e.g. sign changing and non-sign changing $s$-wave states, or
even more interestingly between a triplet and a singlet state, which can be mixed in non-centrosymmetric 
materials; it could also be due to a phase separation, either chemical or a nanoscale electronic
inhomogeneity as in striped phases of cuprates.

Importantly, the present results show that K$_2$Mo$_3$As$_3$ is a non-magnetic electron-phonon 
multi-band superconductor, and based on this the alloy phase diagrams can shed considerable light
on the question of the origin and nature of superconductivity in the K$_2T_3$As$_3$ materials.
The results thus enable sorting out the different proposed superconducting scenarios in this
fascinating family of pnictide superconductors.

\acknowledgements

This work was supported by the U.S. Department of Energy, Office of Science,
Office of Basic Energy Sciences, Award Number DE-SC0019114.
We are grateful for helpful discussions with Keith Taddei.

\bibliography{reference}

\end{document}